\documentclass[pdftex,dvipsnames,usenames,aps,superscriptaddress,twocolumn,amsmath,showkeys]{revtex4-1}

\usepackage{hyperref}
\usepackage{float}
\usepackage{dcolumn}
\usepackage{graphicx}
\usepackage{color}
\usepackage{amssymb}
\usepackage{bold-extra}
\usepackage{caption}
\usepackage{subcaption}

\def\figref#1{Fig.~\ref{fig:#1}}
\def\figlab#1{\label{fig:#1}}  
\def\eqref#1{Eq.~(\ref{eq:#1})}

\newcommand{\Omit}[1]{}

\newcommand{\BLUE}{\color[named]{Blue}}
\newcommand{\Note}[1]{{\BLUE\bf #1}}

\RequirePackage[normalem]{ulem}

\renewcommand{\Note}[1]{{}}

\def\KVI{University of Groningen, KVI Center for Advanced Radiation Technology, Groningen, The Netherlands}
\def\Kapteyn{University of Groningen, Kapteyn Astronomical Institute, Groningen, The Netherlands}
\def\AIVUB{Astrophysical Institute, Vrije Universiteit Brussel, Pleinlaan 2, 1050 Brussels, Belgium}
\def\IIHE{Interuniversity Institute for High-Energy, Vrije Universiteit Brussel, Pleinlaan 2, 1050 Brussels, Belgium}

\def\NIKHEF{Nikhef, Science Park Amsterdam, Amsterdam, The Netherlands}
\def\IMAPP{Department of Astrophysics/IMAPP, Radboud University Nijmegen, Nijmegen, The Netherlands}
\def\CWI{CWI, Centrum Wiskunde \& Informatica, Amsterdam, The Netherlands}
\def\TUe{TU/e, Eindhoven University of Technology, Eindhoven, The Netherlands}
\def\ASTRON{Netherlands Institute for Radio Astronomy (ASTRON), Dwingeloo, The Netherlands}
\def\MPIB{Max-Planck-Institut f\"{u}r Radioastronomie,  Bonn, Germany}

\def\UNH{Department of Physics and Space Science Center (EOS), University of New Hampshire, Durham NH 03824 USA}

\def\Erl{Erlangen Center for Astroparticle Physics, Friedrich-Alexander-Univerist\"{a}t Erlangen-N\"{u}rnberg, Germany}
\def\KIT{Institut f\"{u}r Kernphysik, Karlsruhe Institute of Technology(KIT), P.O. Box 3640, 76021, Karlsruhe, Germany}
\def\DESY{DESY, Platanenallee 6, 15738 Zeuthen, Germany}
\def\CanTho{Department of Physics, School of Education, Can Tho University Campus II, 3/2 Street, Ninh Kieu District, Can Tho City, Vietnam}

\begin{document}

\title{Radio emission from negative lightning leader steps reveals inner meter-scale structure}

\author{B.~M.~Hare} \email[]{B.H.Hare@rug.nl} \affiliation{\KVI}  \affiliation{\Kapteyn}
\author{O.~Scholten} \email[]{O.Scholten@rug.nl} \affiliation{\KVI} \affiliation{\Kapteyn} \affiliation{\IIHE}
\author{J.~Dwyer} \affiliation{\UNH}
\author{U.~Ebert} \affiliation{\CWI} \affiliation{\TUe}
\author{S.~Nijdam} \affiliation{\TUe}
\author{A.~Bonardi}  \affiliation{\IMAPP}
\author{S.~Buitink} \affiliation{\IMAPP} \affiliation{\AIVUB}
\author{A.~Corstanje} \affiliation{\IMAPP} \affiliation{\AIVUB}
\author{H.~Falcke}  \affiliation{\IMAPP} \affiliation{\NIKHEF} \affiliation{\ASTRON} \affiliation{\MPIB}
\author{T.~Huege}  \affiliation{\AIVUB} \affiliation{\KIT}
\author{J.~R.~H\"orandel}  \affiliation{\IMAPP} \affiliation{\AIVUB} \affiliation{\NIKHEF}
\author{G.~K.~Krampah} \affiliation{\AIVUB}
\author{P.~Mitra} \affiliation{\AIVUB}
\author{K.~Mulrey} \affiliation{\AIVUB}
\author{B.~Neijzen} \affiliation{\KVI}
\author{A.~Nelles}  \affiliation{\Erl} \affiliation{\DESY}
\author{H.~Pandya} \affiliation{\AIVUB}
\author{J.~P.~Rachen} \affiliation{\AIVUB}
\author{L.~Rossetto}  \affiliation{\IMAPP}
\author{T.~N.~G.~Trinh}  \affiliation{\CanTho}
\author{S.~ter Veen}  \affiliation{\ASTRON}
\author{T.~Winchen} \affiliation{\AIVUB}

\date{\today}

\begin{abstract}
We use the Low Frequency ARray (LOFAR) to probe the dynamics of the stepping process of negatively-charged plasma channels (negative leaders) in a lightning discharge. We observe that at each step of a leader, multiple pulses of VHF (30~--~80 MHz) radiation are emitted in short-duration bursts ($<10\ \mu$s). This is evidence for streamer formation during corona flashes that occur with each leader step, which has not been observed before in natural lightning and it could help explain X-ray emission from lightning leaders, as X-rays from laboratory leaders tend to be associated with corona flashes. Surprisingly we find that the stepping length is very similar to what was observed near the ground, however with a stepping time that is considerably larger, which as yet is not understood. These results will help to improve lightning propagation models, and eventually lightning protection models.
\end{abstract}


\maketitle

Lightning is one of the most energetic processes in our atmosphere.
It is thought to initiate from a single point, that then separates into positively and negatively charged ends, called positive and negative leaders, which propagate away from the initiation point and into oppositely charged cloud regions~\cite{Dwyer:2014}.
At the tip of each leader many streamer discharges create weakly ionized plasma channels through the joint action of ionization fronts and local field enhancement at the front of the streamer channels.
For positive leaders, electrons accelerate towards the leader, allowing the positive leader to grow fairly gradually while supported by the strong photo-ionization in air as a source of free electrons \cite{Gallimberti2002,  Wormeester:2010, Biagi:2011}. We have recently developed new high-resolution VHF measurement techniques, and applied them to positive leaders \cite{Hare:2019}.

In this work we focus on negative leaders. Negative leaders have a significantly more complex propagation mechanism where they propagate in discrete steps. Each step appears to be due to luminous structures, generally assumed to be conducting (see Ref.~\cite{Malag:2019} for an alternative interpretation) that form in front of the main conducting channel, called space stems in this work. After their formation, these structures grow backward to connect with the main leader body, resulting in a large current pulse to equalize the electric potential.
This process was first observed in laboratory  discharges~\cite{Renardieres81,Gallimberti2002} and later in lightning~\cite{Hill:2011, Biagi:2014, Gamerota:2015}.
However, the majority of the previous work has been done in the optical regime, which does not directly relate to electrical current (e.g. \cite{Malag:2019}), or using radio emission below 10 MHz that is only sensitive larger scale electrical currents (e.g. \cite{Howard:2011, Carlson:2015}). The stepping process has been observed before in VHF emission~\cite{Winn:2011}, however with a resolution that made it difficult to draw firm conclusions.

To investigate the mechanism behind negative leader propagation and its VHF emission we have used LOFAR to provide measurements of the meter-scale distribution of electrical currents in negative leaders using the technique described in Ref.~\cite{Hare:2019}.
These measurements will help to improve lightning leader modeling, which tends to rely on a large number of assumptions, inhibiting, for example, our understanding of basic lightning processes such as attachment to ground, which is critical for improved lightning protection \cite{Cooray:2012,Cooray:2017}. Furthermore, previous work has shown that the majority of terrestrial gamma ray flashes (TGFs), intense bursts of gamma ray radiation with energies up to 10~MeV, are correlated with negative leader stepping \cite{Dwyer:2014}, therefore our improved understanding of leader propagation could be used in future work to help understand TGFs.

We show that each leader step emits a burst of multiple discrete VHF pulses. This is in direct contrast with what is expected based on previous work, which predicts one single VHF source per step \cite{ Carlson:2015}. We find that the majority of VHF sources in a leader step occur within about a meter of each other, showing that VHF radiation from negative leaders comes from corona flashes, which have been observed in laboratory sparks but not in natural lightning \cite{Kochkin:2015, Kostinskiy:2018}. This discovery could explain why lightning leaders tend to emit 100-500~keV X-rays, since similar X-ray bursts seen in laboratory sparks are often associated with corona flashes \cite{ Celestin:2015, Kochkin:2015}.

LOFAR is a distributed radio telescope that is primarily built for radio-astronomy observations~\cite{Haarlem:2013} but has also proven to be an excellent cosmic-ray air-shower detector~\cite{Falcke:2005,Buitink:2016}. Its potential for lightning detection was clear at the initial design~\cite{Holleman:2006}.
We use the Dutch part of LOFAR consisting of thousands of dipole antennas spread over 3200~km$^2$ in the northern Netherlands with antennas operating in the 30~--~80~MHz band.
The traces are sampled at 200~MHz and relative arrival time of each pulse can be measured with about 1~ns accuracy. Our algorithm can locate sources that are at least 120~ns apart.

Previous techniques could map lightning in either 3D with about 100~m accuracy \cite{Rison:1999}, or in 2D with 1$^{\circ}$ accuracy \cite{Stock:2014}. Our technique allows to map lightning in 3D with a horizontal accuracy better than 2~m and 15~m vertically with an efficiency of one source per 1~$\mu$s~\cite{Hare:2019}.

We analyze a lightning flash from September 29$^{th}$ 2017~\cite{Hare:2019}, where the discharge initiated at a height of 4~km. Over time, extended structures are formed with several positive and negative leaders spanning distances of 5~km. In Ref.~\cite{Hare:2019} we focused on the structure of the positive leaders in this flash, and here we analyze the negative ones.
Each located source has an interferometric fit value that lies between 0 and 1, where 1 is the best possible fit. In order to achieve a horizontal location accuracy of 1~m we only used sources that had an interferometric fit value larger than 0.87 while in \cite{Hare:2019} we used sources with fit values larger than 0.85, which resulted in a horizontal location accuracy around 2~m. Our new criterion of 0.87 removes 16\% of the sources that were shown in \cite{Hare:2019}. The qualitative results of this study would not change if we used the lower cut of 0.85 instead of 0.87.

\begin{figure}[h]
	\centering	\includegraphics[width=0.45\textwidth]{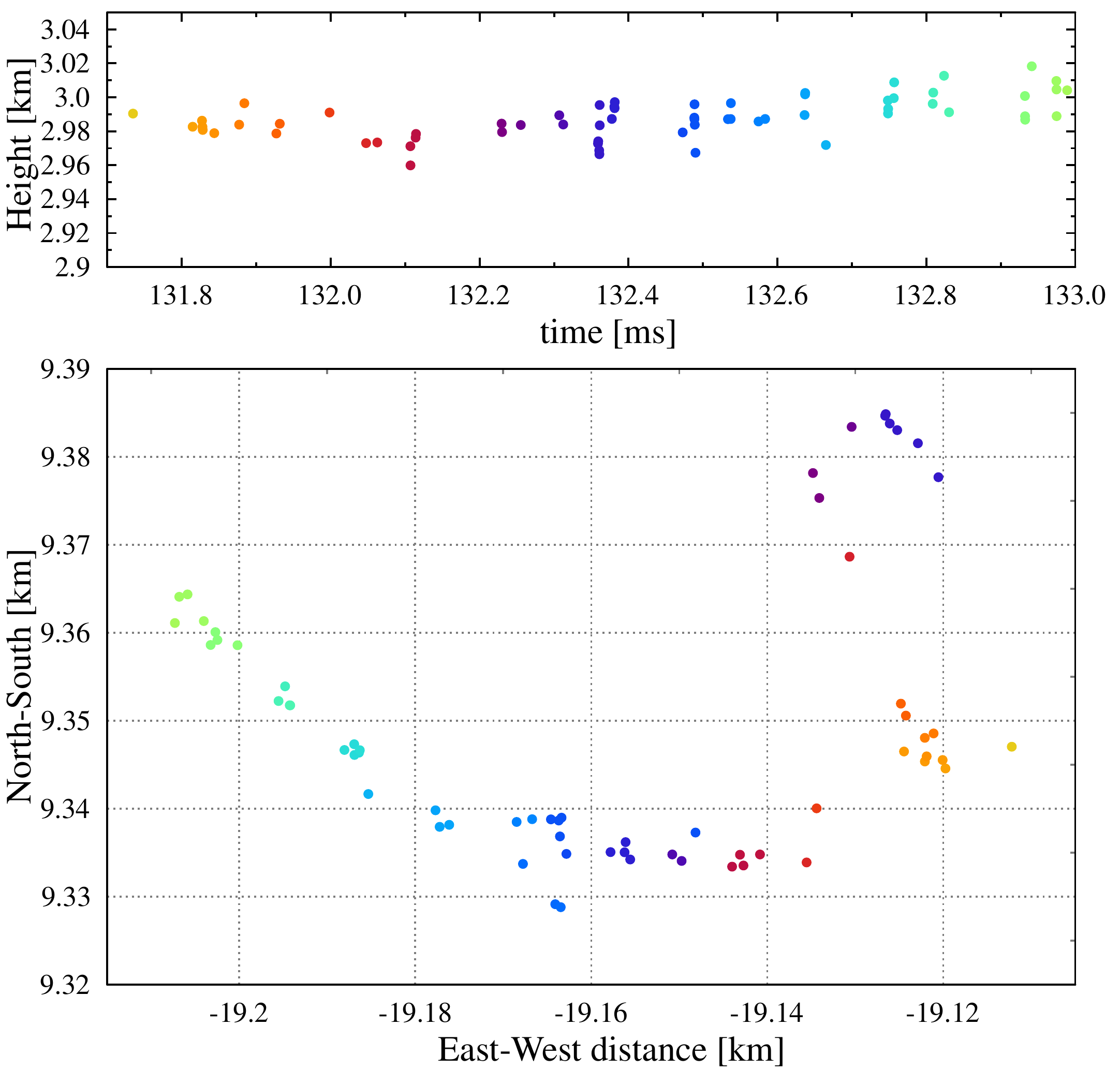}
	\caption{A section of a negative leader where each dot is the location of a reconstructed source. Top panel shows height v.s.\ emission time of the source, the bottom panel shows the projection of the source position on the ground plane where distances are measured from the core of LOFAR. The color of each dot reflects emission time going from yellow to green. }
	\figlab{2017-NegLead-zoom}
\end{figure}

Negative leaders easily extent over distances of several kilometers. The LOFAR measurements allow us to zoom in on a small section of such a negative leader spanning only 100~m as shown in \figref{2017-NegLead-zoom}. Each dot corresponds to the position of a source emitting a single VHF pulse. The location has been determined from the detected arrival time of the pulse at each LOFAR antenna. Each VHF pulse has a full-width half-maximum of about 50~ns, which is mostly dominated by LOFAR's antenna pulse response. We find that these radio sources on negative leaders come in bursts. These bursts can be seen in \figref{2017-NegLead-zoom}, where the VHF sources tend to cluster in time. A wider view of this leader is shown in the supplementary materials. These bursts are observed across all well-imaged negative leaders. The obvious interpretation is that these bursts are due to leader stepping.

\begin{figure}[ht]
	\centering
	\includegraphics[width=0.47\textwidth]{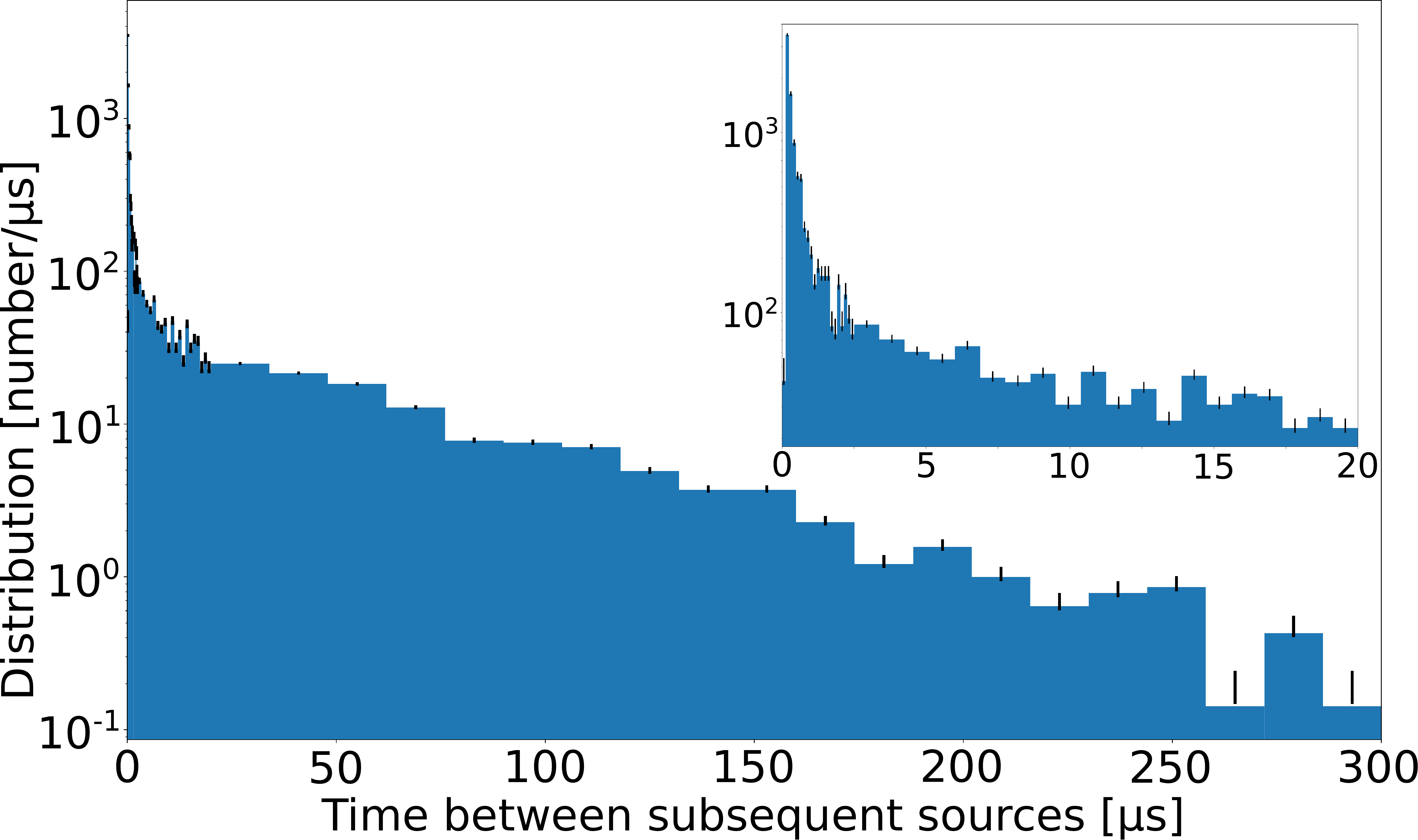}   
	\caption{The distribution of time between subsequent sources. The error bars show the lowest and largest poisson rates that can model the data with 34\% confidence (1~$\sigma$). }
	\figlab{time_dist}
\end{figure}

\figref{time_dist} shows a distribution of time between subsequent radio sources across 26 negative leader segments, with a total length of about 15 km, in the 2017 lightning flash. If every radio source were randomly distributed in time, the time between radio sources would be exponentially distributed. The distribution strongly deviates from an exponential, in particular, there is a sharp spike below time differences of 4~$\mu$s that shows that our located sources cluster together in time significantly more than could be possible for random chance. This spike continues down to 120~ns, the smallest time differences that can be probed with our present imaging algorithm.

It is ambiguous how to precisely define which sources should be clustered together in a burst. This is expressed by the fact that the distribution in \figref{time_dist} is very smooth. In lieu of a physics-inspired definition, we have defined a burst such that every located source in a burst is within 2~$\mu$s of its subsequent radio source. This time cut was chosen because: 1) it includes the majority of VHF sources shown in the spike in \figref{time_dist}, 2) it is short enough that it minimizes the chance of VHF sources from different leader steps to interfere with our results, and 3) the qualitative results are similar even if the time is halved or doubled. The number of sources in a burst may thus vary from a single one up to a maximum of 9 located sources, using this prescription.
Of the total of 2599 bursts we have 224 bursts with 3 or more sources. Investigation of the VHF time traces shows that the majority of bursts with a single located source even have multiple VHF pulses that are not located but most probably come from the same spot.
If we use instead 8~$\mu$s in the definition of a burst we obtain qualitatively very similar numbers (2204 bursts of which 340 have 3 or more sources).
We find that the strength of the pulses within a burst varies greatly, in addition some burst may contain three strong pulses while others may have a single much weaker one.

\begin{figure}[ht]
	\centering
	\includegraphics[width=0.47\textwidth]{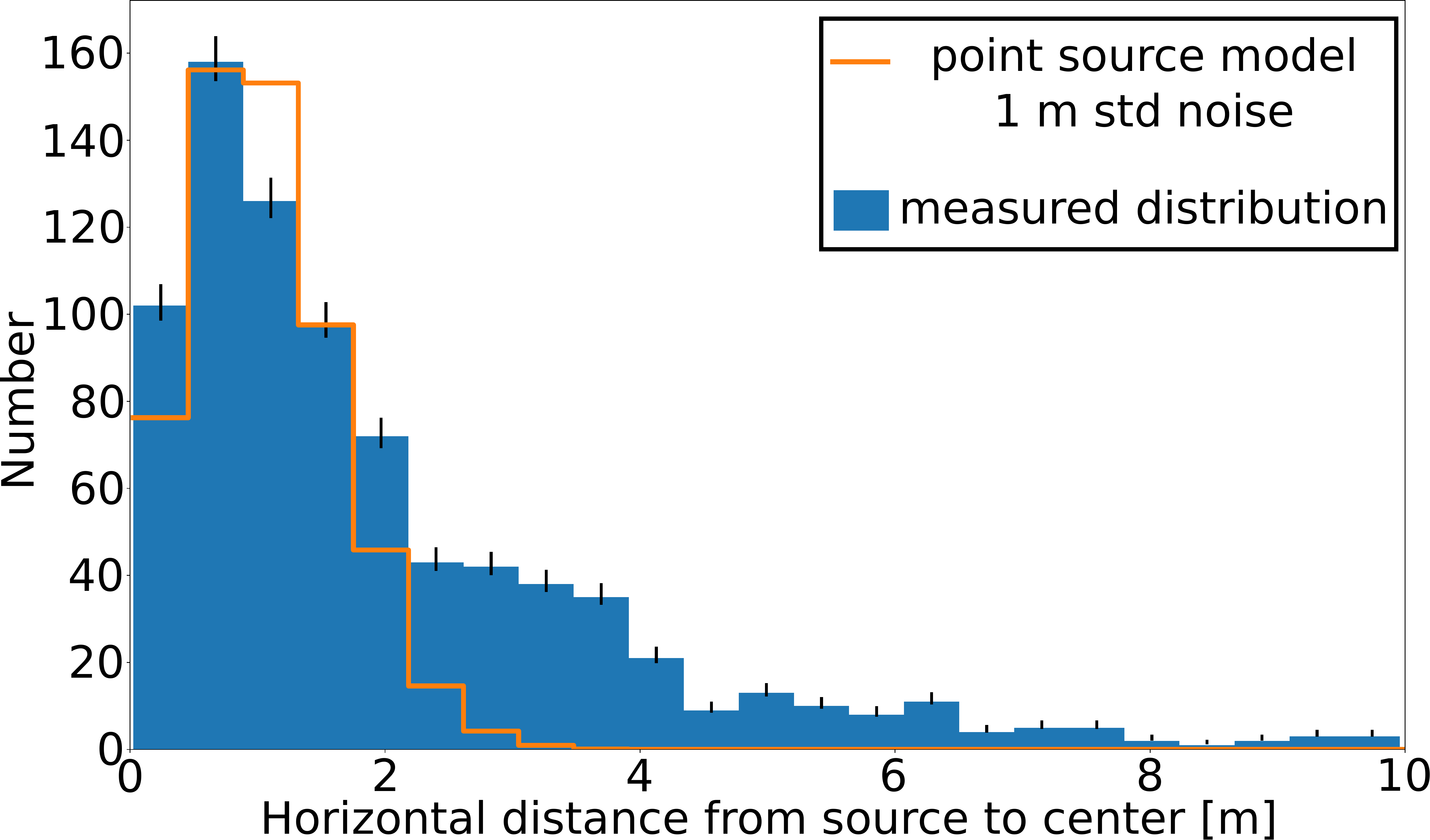}   
	\caption{The histogram shows the horizontal spatial distribution of pulses in a bursts. The orange line gives the simulation results if the sources are at one location with a 1~m horizontal location accuracy and accounts for 2/3 of the number of sources. The error bars indicate the lowest and highest poisson rates that could model the data within 34\% confidence (1~$\sigma$).}
	\figlab{spatial_dist}
\end{figure}

\figref{spatial_dist} shows the spatial distribution of sources within a burst that have three or more located sources by binning the horizontal distance between each source and the geometric center of all pulses in the burst. Note that we focus on the horizontal plane, since our horizontal location accuracy (around 1~m) is significantly better than our vertical location accuracy (around 10~m). Also shown is a simulated distribution if every radio source in a burst came from the same location with a location error of 1~m. The fact that 2/3 of the data-derived distribution is within the radius of our simulation shows that our data is consistent with the majority of located sources in a burst coming from the same location. This can also be seen from \figref{2017-NegLead-zoom} where the different bursts in the time v.s.\ height plot are also localized after projecting on the ground plane. However there are also many bursts where the sources are spread over larger distances. This is expressed by the shoulder in the distribution at a distance of 3.5~m. This shoulder persists independent of changes in our quality cut, burst definition\Note{ see \figref{spatial_dist-8}}, or specific set of leaders used in the analysis.
Bursts with spatial extent seem to be mostly due to simultaneous activity in close branches, and due to bursts that are extended length-wise along the channel (with lengths around 5~m). The supplementary material includes figures that show a variety of different bursts.
\Note{ see red sources at t=128.6 and t=128.8 in \figref{2017-NegLead-anotherzoom} and see dark blue sources at t=192.2 in \figref{2017-NegLead-anotherzoom}}

The total duration of a burst (for bursts with at least 2 pulses) is exponentially distributed with a median of 0.5~$\mu$s and a suppression below 0.1~$\mu$s. Changing our burst definition to 8~$\mu$s increases the median considerably to 1.5~$\mu$s by adding a long tail extending to 4~$\mu$s. \Note{see \figref{duration}}
Even though the density of located sources in the flash is second to none, it should be realized that our imaging formalism has an efficiency of only 30\%, i.e.\ only a third of the strongest pulses in a spectrum is located.  This probably most strongly affects the burst duration.
For example, if a pulse in the middle of a burst is not imaged, then our simple 2~$\mu$s definition may split that burst in two. While, using a 8~$\mu$s definition, may combine multiple bursts.

\begin{figure}[h]
	\centering{
	\includegraphics[width=0.47\textwidth]{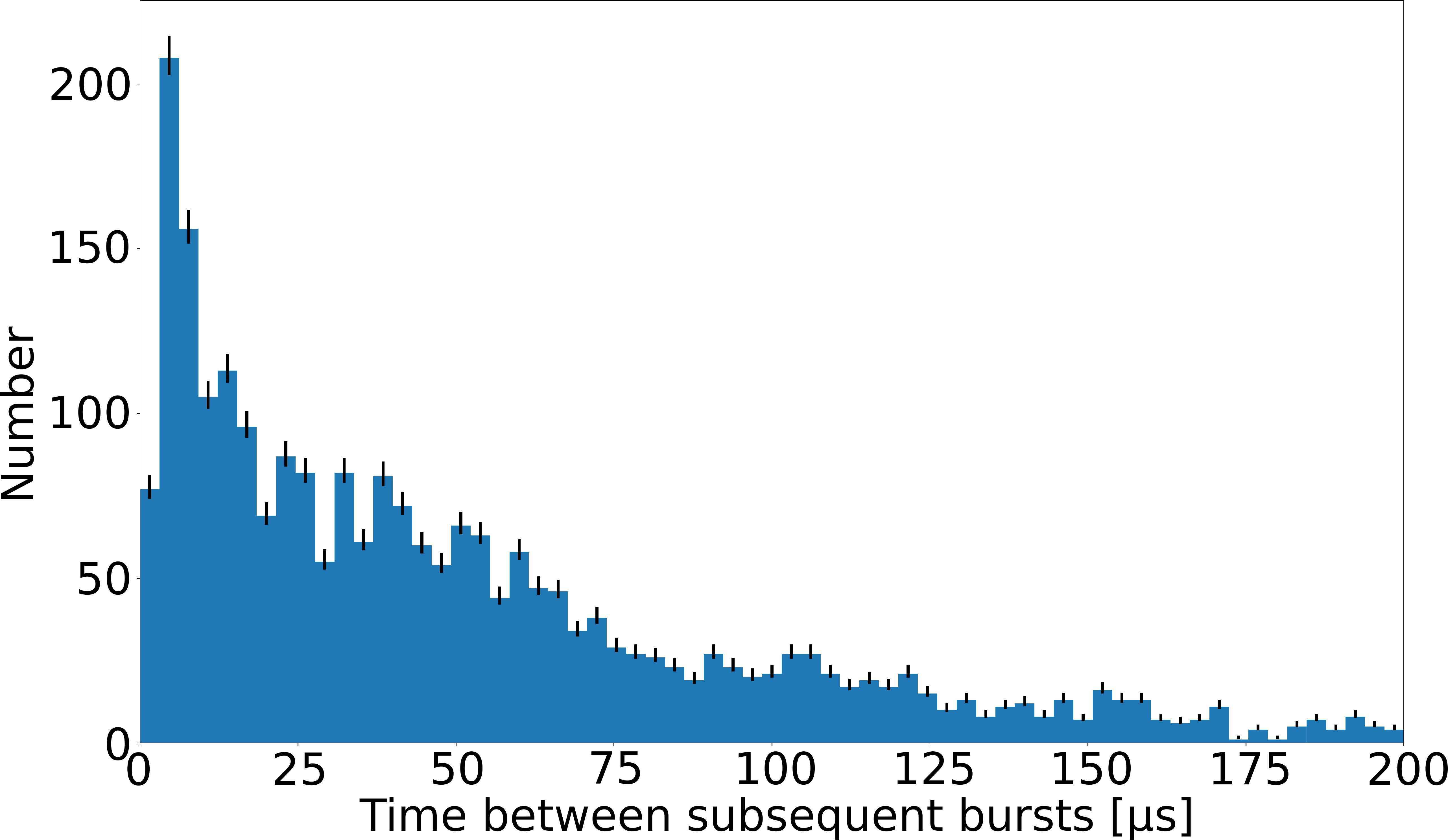}   
	\includegraphics[width=0.47\textwidth]{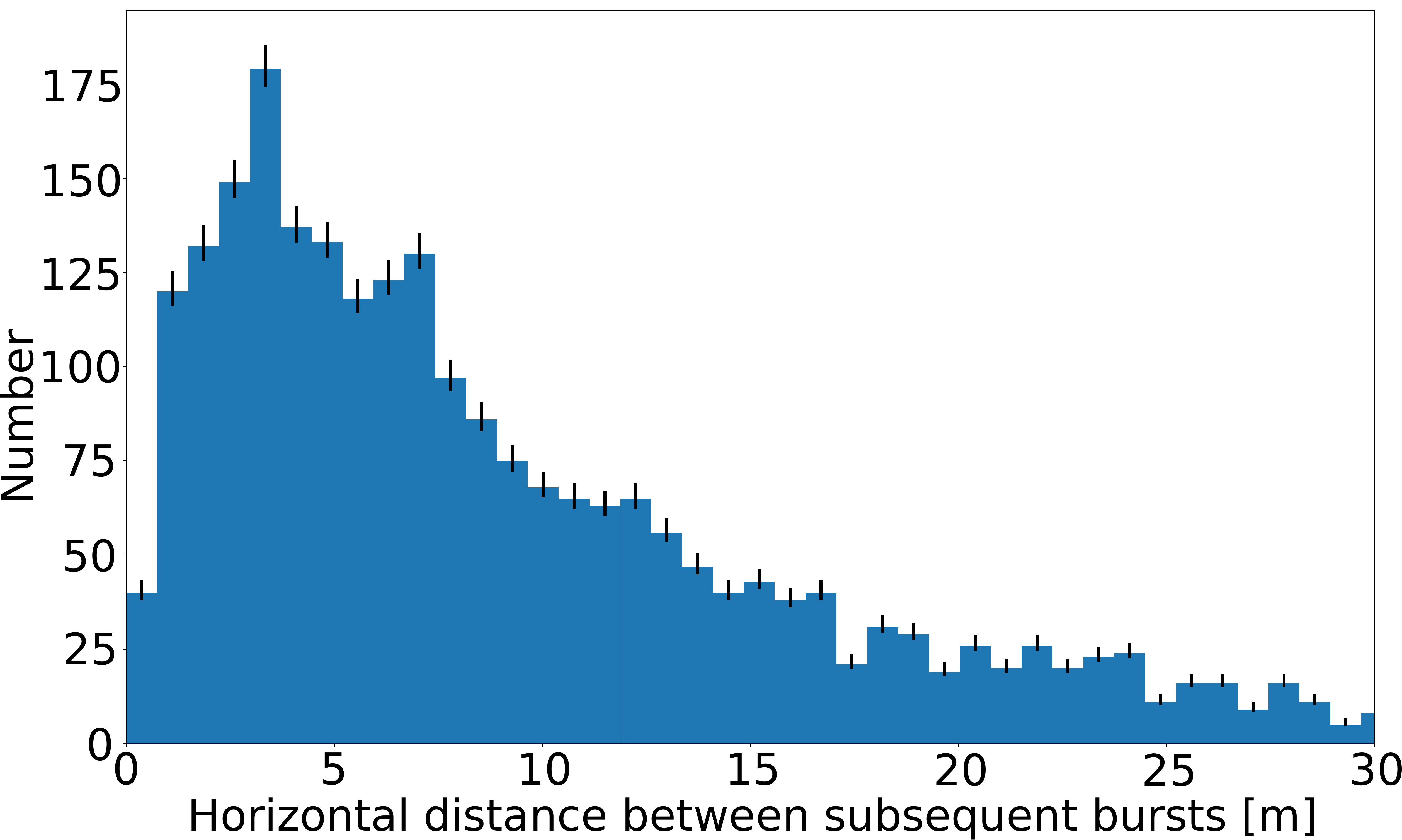} }  
	\caption{Time between bursts (top panel) and horizontal distance between bursts (bottom panel), where a burst can have only a single pulse.}
	\figlab{Between}
\end{figure}

The time distribution between bursts shows an exponential distribution, see \figref{Between}, with a median at 40~$\mu$s which we interpret as the median stepping time. This number is not affected much by the precise definition of a burst since using the 8~$\mu$s burst definition yields a median stepping time of 50~$\mu$s. The horizontal distance between bursts also shows an exponential distribution with a cut-off at distances below 4~m and a median of 7.5~m where the median shifts to 8.5~m for the 8~$\mu$s burst definition. We have taken here the distance as measured in the horizontal plane because our vertical resolution is of the order of 10~m and thus would confuse the picture.
Optical observations of leader growth well below the cloud have found that the time between negative leader steps tends to be around 10~$\mu$s and their length tends to be around 5~m~\cite{Hill:2011}.
We thus observe that in the cloud the stepping time is considerably longer than close to ground with a stepping length that is only marginally larger. Near ground level one could expect stronger electric fields than in a cloud but how this reflects in the observed differences is not understood.

A very common interpretation of the negative-leader stepping process is that when the space stem connects to the existing leader, there is a current pulse that equalizes the voltage in the space stem to the much larger (negative) voltage at the tip of the leader by removing positive charge from the space stem \cite{ Carlson:2015}.
In the process, the space stem, which was poorly conducting, heats up due to the dissipation of the electric energy of the current, ionizes further and becomes a good conducting heated plasma thus forming a new leader section.
This sudden voltage jump will cause the electric field to exceed breakdown ($E_c=3.2$~MV/m in air at STP) and produce an inception cloud, where an semi-spherical ionization front expands until the electric field on its surface drops to the breakdown value. Subsequently this ionization front breaks-up into many streamers~\cite{Chen:2015,Kostinskiy:2018}, thin channels of ionized air that grow due to the field enhancement and ionization fronts at their tips. The process of suddenly forming a multitude of streamers during a leader step is also known as a corona flash~\cite{Bazelyan_CoronaFlash:2000, Kochkin:2014}.
The streamers will then, somehow, generate a new space stem that will then grow backwards towards the leader, allowing the whole process to repeat. It is thought that a space stem is formed from merging streamers, but the process is not understood. An alternative theory was recently suggested in Ref.~\cite{Malag:2019}.

One could imagine that the radio emission is due to the large current pulse at the time when the space stem connects to the main leader channel, however, this simple model is not consistent with our data since it predicts a single pulse at each leader step while we observe a whole burst of pulses. This single pulse is, however, regularly observed below 10~MHz \cite{Biagi:2014, Howard:2011}. In Ref.~\cite{Shi:2018} a model is proposed where VHF emission is generated by collisions between streamers, but it is not clear how to compare this model to our data.

A likely mechanism for the emission of pulses in bursts is the process where the inception cloud breaks up into a multitude of streamers~\cite{Briels:2008}, much like has been observed in laboratory experiments~\cite{Briels:2008, Chen:2015,Kostinskiy:2018}.
This fits our observation in \figref{spatial_dist}, that the dominating emission is nearly point-source like at the tip of the leader, with a few sources coming from a short distance ($\approx$~3.5~m) along the body of the leader.
Note that positive leaders often do not exhibit corona flashes, which could explain why negative leaders emit significantly more VHF radiation than positive leaders~\cite{Kochkin:2012}.

Based on the amplitude of the pulses we observe, we infer that energy emitted by the strongest radio source regions that we receive has an order-of-magnitude of $4\times10^{-6}$~J in our 30-80~MHz frequency band. This roughly equates to a streamer with an order-of-magnitude of $5\times10^{13}$ free electrons. Details of these order-of-magnitude calculations are given in Supplemental Material, which includes Refs.~\cite{Neijzen:2019,Mulrey:2019,Qin:2014,Briels:2006,Luque:2014}. This is consistent with the idea that there are of the order of $10^5$ steamers \cite{Celestin:2015} in a corona flash, distributed in strength of emitted VHF energy, where we are only sensitive to the extreme tail of that distribution. Future work is needed to find the distribution of detected streamer sizes.

As mentioned before, the large current pulse during a step moves the negative charge cloud over the length of the step. The radio emission during this step must have a wavelength of at least the spatial extent of the charge cloud (expected to be 10's of meters) to be coherent and thus strong.
Thus, the radiation from the stepping current itself has a peak intensity at frequencies well below the LOFAR band of 30~--~80~MHz (10~--~3.8~m) which would explain why this signal is not clearly visible in our data.
It therefore would be very interesting to perform simultaneous measurements in the 100~kHz~--~10~MHz band, where such current pulses are regularly observed.



In this work we have established that the VHF emission seen from stepping negative leaders in lightning are most likely due to streamer formation around the region of the step.
The VHF emission appears concentrated near the tip of the leader, potentially where the inception cloud breaks up into streamers during a corona flash, which has not been observed in natural lightning before. There is also emission along the body of the step, potentially due to spurious streamer emission from the body of the leader.

\begin{acknowledgements}
The LOFAR cosmic ray key science project acknowledges funding from an Advanced Grant of the European Research Council (FP/2007-2013) / ERC Grant Agreement n. 227610. The project has also received funding from the European Research Council (ERC) under the European Union's Horizon 2020 research and innovation programme (grant agreement No 640130).   We furthermore acknowledge financial support from FOM, (FOM-project 12PR304). AN is supported by the DFG (NE 2031/2-1). TW is supported by DFG grant 4946/1-1.
\\
This paper is based on data obtained with the International LOFAR Telescope (ILT). LOFAR~\cite{Haarlem:2013} is the Low Frequency Array designed and constructed by ASTRON. It has observing, data processing, and data storage facilities in several countries, that are owned by various parties (each with their own funding sources), and that are collectively operated by the ILT foundation under a joint scientific policy. The ILT resources have benefitted from the following recent major funding sources: CNRS-INSU, Observatoire de Paris and Universit\'{e} d'Orl\'{e}ans, France; BMBF, MIWF-NRW, MPG, Germany; Science Foundation Ireland (SFI), Department of Business, Enterprise and Innovation (DBEI), Ireland; NWO, The Netherlands; The Science and Technology Facilities Council, UK.
\end{acknowledgements}

\bibliography{NegativeLeaders-v20_sub}

\end{document}